\documentclass[twocolumn,aps,prb,showpacs]{revtex4-1}
\usepackage{graphicx}
\usepackage{amssymb}
\usepackage{bm}
\usepackage{amsmath}

\makeatletter

\begin{document}

\title{Optical signatures of the tunable band gap and valley-spin coupling
 in silicene}
\author{L. Stille$^1$}
\author{C.J. Tabert$^{1,2}$}
\author{E.J. Nicol$^{1,2}$}
\email{enicol@uoguelph.ca}
\affiliation{$^1$Department of Physics, University of Guelph,
Guelph, Ontario N1G 2W1, Canada} 
\affiliation{$^2$Guelph-Waterloo Physics Institute, University of Guelph, Guelph, Ontario N1G 2W1 Canada}

\date{\today}
\begin{abstract}
We investigate the optical response of the silicene and similar materials,
such as germanene, in the
presence of an electrically tunable band gap for variable doping. 
The interplay of spin orbit
coupling, due to the buckled structure of these materials, and a perpendicular
electric field gives rise to a rich variety of phases: a topological or
quantum spin Hall insulator,
a valley-spin-polarized metal and a band insulator. We show that the
dynamical conductivity should reveal signatures of these different phases
which would allow for their identification along with the determination of
parameters such as the spin orbit energy gap. We find an interesting 
feature where
the electric field tuning of the band gap might be used
to switch on and off  the Drude intraband response.  
Furthermore, in the presence
of spin-valley coupling, the response to circularly polarized light as a 
function of frequency and electric field tuning of the band gap is examined.
Using right- and left-handed circular polarization it is possible to
select a particular combination of 
spin and valley index. The frequency for this effect
can be varied by tuning the band gap.
\end{abstract}


%
\maketitle

\section{Introduction}

\begin{figure}
  \begin{center}
    \includegraphics[width=0.9\linewidth]{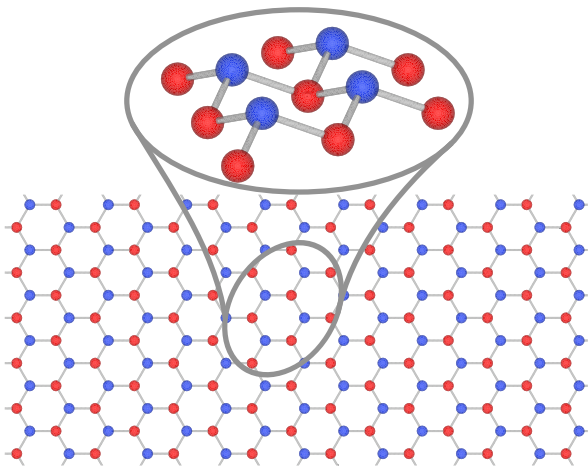} 
  \end{center}
  \caption{\label{fig1}(Color online) 
The crystal structure of silicene is based on the honeycomb lattice
shown here, however, due to the size of the silicon atom,
the the honeycomb lattice is buckled with the A and B sublattices
shifted out of a common two-dimensional plane. [Crystal structure
plotted using VESTA\cite{Momma:2011}.]
}
\end{figure}

The experimental realization of two-dimensional crystals has sparked a flurry of
research activity. The prototypical example was the isolation of 
graphene,\cite{Novoselov:2004,Novoselov:2005a} a single layer of carbon atoms,
 with its remarkable physics due to the Dirac fermionic
nature of its charge carriers. Much research has ensued\cite{Neto:2009,Abergel:2010,DasSarma:2011,Kotov:2010} and part of the
focus is on the potential to manipulate the physics of these charge carriers
to benefit technological applications. In graphene the charge carriers show
a linear dispersion near the Fermi level\cite{Wallace:1947} 
with chiral properties attributed to
the topological aspect of two sublattices in a hexagonal crystal lattice. This
leads to a mapping of the low energy nearest neighbor tight-binding 
Hamiltonian on to a Dirac equation for
massless fermions which exist at $K$ and $K'\equiv -K$ 
in the hexagonal Brillouin zone.\cite{Semenoff:1984} The conical disperison at these $K$ points
are referred to as valleys and many properties in graphene are spin degenerate
and valley degenerate. As the spin-orbit interaction in  graphene is quite 
small\cite{Konschuh:2010} and therefore neglected, 
interesting effects, such as inducing
 spin-valley coupling discussed here, will not be evident. 
This is unfortunate as much interesting physics and potential
for spin manipulation will not be manifest in graphene.
Given some of the limitations of
graphene in this regard, attention has turned to other 2D crystals which
have graphene-like features but stronger spin-orbit coupling.
 One such material is silicene, a monolayer
of silicon atoms arranged in a honeycomb lattice, as in graphene. As silicon
is already used in electronics and a large silicon manufacturing
 industry is in place,
it bears closer inspection. Silicene has only recently been made in ribbons\cite{Aufray:2010,Padova:2010,Lalmi:2010}
and an experimental literature has yet to be developed. However,
numerical calculations have shown that it should be possible to make silicene
suitable for transfer to a substrate where it might be 
gated.\cite{Cahangirov:2009,Drummond:2012}
As silicon has a stronger spin-orbit coupling than graphene, it will
exhibit more clearly an energy gap
in the band structure. It has been argued that
in this circumstance, the Dirac nature of the fermions makes
a topological insulator (TI). One of the
interesting predictions for silicene involves that of inducing
a  tunable band gap\cite{Drummond:2012} which would give rise to a transition between a topological
insulator and a band insulator (BI).\cite{Drummond:2012,Ezawa:2012njp} This comes about due to the fact that in
silicene, the size and distance between silicon atoms leads to a buckling
of the structure where one sublattice is  shifted upwards out of the
2D plane relative to the other (see Fig.~\ref{fig1}). 
This means that an applied electric
field perpendicular to this plane changes the sublattice potential and produces a 
tuning of the band gap in the dispersion.\cite{Drummond:2012} 
Furthermore, the bands are spin-split
in a valley-dependent way allowing for an examination of spin-valley coupling.\cite{Ezawa:2012opt}
Other examples of systems discussed in the literature which are predicted to
show spin-valley coupling are MoS$_2$ and other
group-VI dichalcogenides\cite{Xiao:2012,Cao:2012} and a monolayer
 of germanium (germanene)\cite{Liu:2011}. Currently,
MoS$_2$ has been the subject of several experimental investigations, for example Ref.~\cite{Mak:2010,Mak:2012,Zeng:2012,Cao:2012,Sallen:2012}, 
while silicene is still in its infancy and germanene has yet to be
synthesized. Nonetheless, we will use the examples of silicene and
germanene
as a models to
examine issues associated with spin-valley coupling and optical signatures
of a TI to BI transition.

In this work, we study the finite frequency conductivity of a material
such as silicene which can be induced to
have spin-valley coupling and a tunable band gap.
We find that there are characteristics in the
far infrared or THz conductivity which would allow for the
differentiation of the TI versus the BI. With tuning of the band gap, these
features could be used to identify the critical point of transition and allow
for an experimental determination of the strength of spin orbit coupling in
these systems. We also consider the effect of finite doping and 
circularly polarized
light as possible probes of spin-valley coupling. Previous issues associated
with circular dichroism are revisited in view of performing 
broadband spectroscopy.

Our paper is organized in the following manner. In Section~II, we briefly
describe the basic Hamiltonian and approximations used in this work and
outline the nature of our calculation of finite frequency conductivity. Following this,
in Section~III, we feature our results for the absorptive part of the
longitudinal
optical conductivity and
provide a discussion of the physics. In Section~IV, we discuss the case of
circularly polarized light which illustrates the potential
 for  spin-valley polarization and its the tuning 
by electric field. Our conclusions are summarized in
Section~V.

\begin{figure}
  \begin{center} 
      \includegraphics[width=0.9\linewidth]{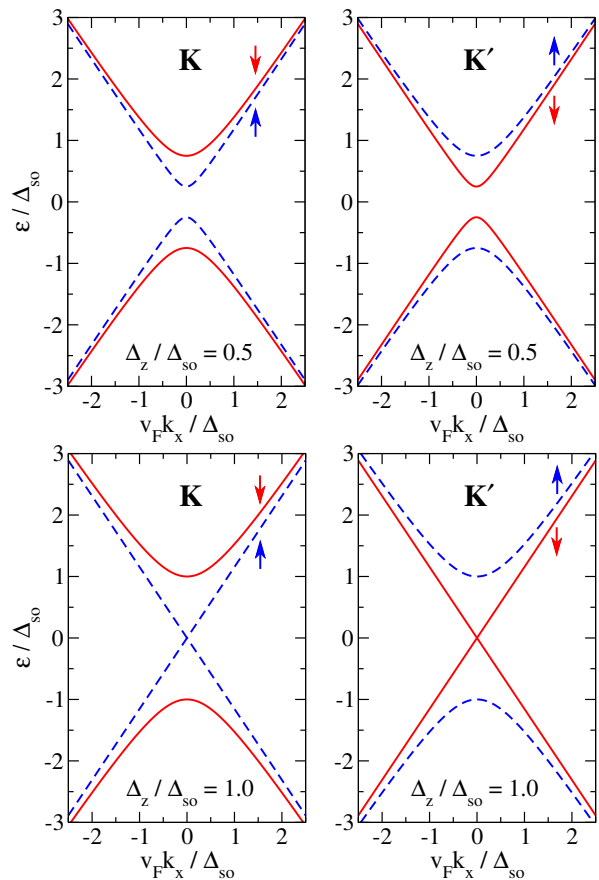} 
  \end{center}
  \caption{\label{fig2}(Color online) The presence of 
spin orbit coupling and
a perpendicular electric field, parametrized by $\Delta_{\rm SO}$
and $\Delta_z=E_zd$, respectively, gives rise to spin-split bands 
about the $K$ point, with two gaps as shown in
the upper left, one of which may
be tuned to zero at $\Delta_z=\Delta_{\rm SO}$ as shown in the lower left.
The bands at $K'$ are reversed from those at $K$ (right frames).
}
\end{figure}

\section{Theoretical Background}

\begin{figure}
  \begin{center} 
      \includegraphics[width=0.9\linewidth]{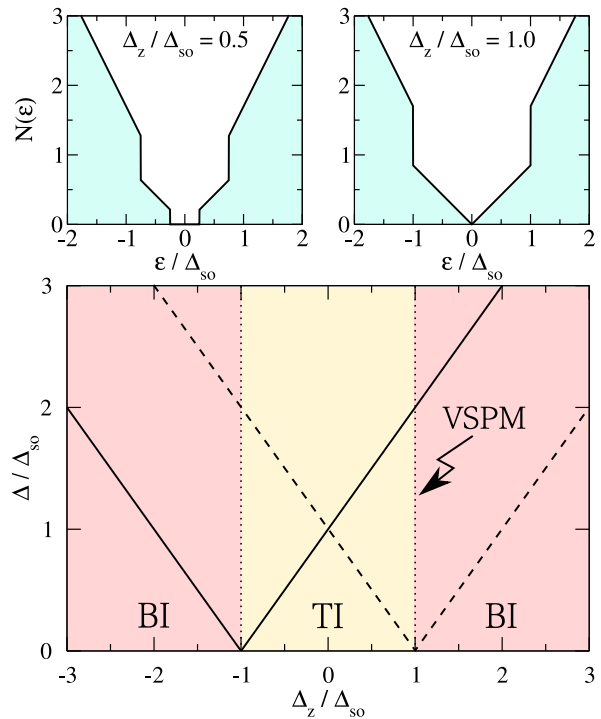} 
  \end{center}
  \caption{\label{fig3}(Color online) 
Upper frames: The total density of states for the band structures
illustrated in Fig.~\ref{fig2}. The right frame is the case for
the VSPM state. The variation with $E_z$ of the two edges shown in the left
frame will follow the trend shown in the lower frame. Lower frame:
The evolution of gap in the spin down (solid curve) and spin up (dashed curve)
band about the $K$ point as a function of $\Delta_z=E_zd$. The regions
of band insulator (BI) versus the topological insulator (TI) are indicated
by shading and the dotted line marks the valley-spin-polarized metal
(VSPM) state. 
}
\end{figure}

Recently, there have been theoretical works 
which have calculated the band structure of
silicene\cite{Liu:2011,Drummond:2012} using density functional theory
 including effects due to spin-orbit coupling (SOC) and a perpendicular
electric field. The spin-orbit band gap has been found to be about
1.5 meV and this can be increased to 2.9 meV under strain.\cite{Liu:2011}
This latter study has also predicted a SOC 
band gap of 23.9 meV for  2D germanium or germanene. 
In the work of Drummond et al\cite{Drummond:2012}, a SOC gap of 1.4-1.5 meV is also found for silicene and  with a  perpendicular
electric field, the resulting gap  may be tuned up to tens of meV before
other transformations occur. This work along with
 that of Ezawa\cite{Ezawa:2012njp,Ezawa:2012prl,Ezawa:2012opt} suggest that
much of this behavior may be captured by a simple low energy 
Hamiltonian written about the K point labeled $K_{\xi}$:
\begin{equation}
\hat H_{K{_\xi}}=v(\xi k_x\hat\tau_x+k_y\hat\tau_y)-\xi\frac{1}{2}
\Delta_{\rm SO}\hat\sigma_z\hat\tau_z+\frac{1}{2}\Delta_z\hat\tau_z,
\label{eq:Hpm}
\end{equation}
where $\xi=\pm$ is the valley label for the two K-points,
$K_\pm$ which were previously referred to as $K$ and $K'$. 
The first term is the Hamiltonian for Dirac electrons with
velocity $v$ (we take $\hbar=1$), the second
term is the Kane-Mele SOC term with SOC gap $\Delta_{\rm SO}$
 and the last term is associated with
the perpendicular 
electric field $E_z$ which, when applied across the buckled structure, gives
rise to an A-B sublattice asymmetry with on-site potential $\Delta_z=E_zd$
where $d$ is the perpendicular distance between the two sublattice planes.
The matrices
$\hat\tau_i$ and $\hat\sigma_i$ are the
Pauli
matrices. The former act on the pseudospin space
associated with the A and B sublattices and the latter are for
real spin with $\hat\sigma_z$, the z-component. 
In the papers by Ezawa, a Rashba SOC is included in addition to the intrinsic SOC used here. As the Rashba SOC is an order of 10
smaller in magnitude, we neglect it.  Therefore, the full 8x8 matrix spanning both $K$ points is block diagonal in
2x2 matrices labeled by valley index $\xi=\pm$ for
$K$ and $K'$ and spin
index $\sigma=\pm$ for spin up and spin down, respectively: 
\begin{equation}
\hat H_{\sigma\xi}=\left(\begin{array}{cc}
-\frac{1}{2}\sigma\xi\Delta_{\rm SO}+\frac{1}{2}\Delta_z & v (\xi k_x-ik_y)\\
v (\xi k_x+ik_y) & \frac{1}{2}\sigma\xi\Delta_{\rm SO}-\frac{1}{2}\Delta_z\end{array}\right).
\label{eq:Hmatrix}
\end{equation}
The eigenvalues are evaluated from the Hamiltonian above 
to be:
\begin{equation}
\epsilon_{\sigma\xi}=\pm\sqrt{v^2k^2+\frac{1}{4}(\Delta_z-\sigma\xi\Delta_{\rm SO})^2}.
\label{eq:eigen}
\end{equation}
These are plotted in Fig.~\ref{fig2} for a value of $\Delta_z=0.5\Delta_{\rm SO}$
where the bands are spin-split, reversed at the other $K$ point, and represent
a topological insulator. Also shown is the case where $\Delta_z=\Delta_{\rm SO}$.
At this critical point, the gap of one of the spin-split bands closes to give
a Dirac point while at the other $K$ point it is gapped and it is the other
spin band which has no gap. This has been
termed a valley-spin-polarized metal (VSPM).\cite{Ezawa:2012prl}
For $\Delta_{z}>\Delta_{\rm SO}$, the
spectrum becomes fully gapped again but the system is 
a band insulator albeit with unusual chiral properties.\cite{Drummond:2012}
In the case of $\Delta_z=0$, the bands are spin degenerate with a single
band gap of $\Delta_{\rm SO}$.
In Fig.~\ref{fig3}, we show the total density of electronic states for the
bands shown in Fig.~\ref{fig2}.
The density of states reflects the two gaps that open in the system
depending on the strength of $E_z$. At the critical value 
where $\Delta_z=\Delta_{\rm SO}$, which is the VSPM state,
 the low energy density of states
is linear out of zero energy and jumps at a value equal to the sum of
$\Delta_z+\Delta_{\rm SO}$ to further linear behavior with double the
slope (upper righthand frame of Fig.~\ref{fig3}). 
This jump reflects the spin-split bands. 
For finite $\Delta_z\ne\Delta_{\rm SO}$, a low energy gap appears
at a value of the difference of these two energies and a jump
at higher energy remains (upper lefthand frame). 
If $E_z=0$, there is
only the spin-orbit gap and the spin degeneracy in this case
gives just the one gap edge with no further jumps in the density of
states. This behavior is embodied in the following formula for the
density of states:
\begin{equation}
N(\epsilon)= \sum_{\sigma,\xi} N_{\sigma\xi}(\epsilon)
\end{equation}
where
\begin{equation}
N_{\sigma\xi}(\epsilon)=\frac{|\epsilon|}{2\pi\hbar^2v^2}\Theta(2|\epsilon|-|\Delta_{\sigma\xi}|),
\end{equation}
and $\Delta_{\sigma\xi}\equiv\Delta_z-\sigma\xi\Delta_{\rm SO}$, with $\sigma=\pm$ indexing the spin up/down bands at the $K_\xi$ point.
A plot of the energy gaps for the spin-resolved bands are shown in 
the lower frame of Fig.~\ref{fig3}, where 
the solid curve follows the notation of Fig.~\ref{fig2} and indicates
the gap for the spin down band and the dashed line
is for the spin up band at $K$. The solid and dashed lines would reverse 
in the plot for
the $K'$ point, as seen in Fig.~\ref{fig2}. The region of electric field
that gives $|\Delta_z|<\Delta_{\rm SO}$ has been shown to pertain to a 
TI 
of the Kane-Mele type (or a quantum Hall spin insulator)
and for $|\Delta_z|>\Delta_{\rm SO}$, the system is a BI. The
critical point at which $|\Delta_z|=\Delta_{\rm SO}$ is the VSPM state
discussed previously. It is clear from this figure that in the BI state,
both gap edges in the DOS will increase with the magnitude  of
the perpendicular 
electric field, whereas in the TI case, the higher
one will
increase with electric field while the lower one decreases, which would be
a method of differentiating between the density of states of the BI versus the TI.

Such considerations also will apply to the dynamical
conductivity which we evaluate by
the usual Kubo formalism.\cite{Nicol:2008,Carbotte:2010,Tabert:2012}
The zero temperature, absorptive part of the conductivity at photon frequency
$\Omega$ is given from an evaluation of 
\begin{align}
\sigma_{ij}(\Omega)&=\frac{e^2}{2\Omega}\int_{|\mu|-\Omega}^{|\mu|}\frac{d\omega}{2\pi}
\int\frac{d^2 k}{(2\pi)^2}\,\text{Tr}\left[\hat{v}_i\hat{A}({\bm k},\omega+\Omega)\hat{v}_j\hat{A}({\bm k},\omega)\right],
\label{eq:cond}
\end{align} 
where $\mu$ is the chemical potential, 
$\hat v_i=\partial \hat H/\partial k_i$, with $i,j$ referring to
spatial coordinates in 2D, and
$\hat A({\bm k},\omega)$ is the spectral function of the Green's function found
from $\hat{G}^{-1}(z)=z\hat{I}-\hat{H}$ through the relation
\begin{equation}
\hat{G}_{ij}(z)=\int_{-\infty}^{\infty}\frac{d\omega}{2\pi}\frac{\hat{A}_{ij}(\omega)}{z-\omega}.
\end{equation}
Using Eq.~\eqref{eq:cond} and the appropriate $v_i$ operator, we obtain the expression for the real part of the zero temperature longitudinal conductivity
\begin{equation}
\text{Re}\sigma_{xx}(\Omega)=\sum_{\sigma,\xi}\text{Re}\sigma^{\sigma\xi}_{xx}(\Omega),
\label{eq:totcondxx}
\end{equation}
where the response associated with a particular spin in a particular valley
is given as
\begin{align}
\text{Re}&\sigma_{xx}^{\sigma\xi}(\Omega)
=\frac{e^2}{2\Omega}\int_{|\mu|-\Omega}^{|\mu|}\frac{d\omega}{2\pi}\int\frac{d^2k}{(2\pi)^2}v^2
\notag\\
&\times
\bigg[
A^{\sigma\xi}_{11}(\bm{k},\omega)A^{\sigma\xi}_{22}(\bm{k},\omega+\Omega)
+A^{\sigma\xi}_{22}(\bm{k},\omega)A^{\sigma\xi}_{11}(\bm{k},\omega+\Omega)
\bigg],
\label{eq:numcondxx}
\end{align}
with
\begin{equation}\label{eq:A11}
A^{\sigma\xi}_{11}(\bm{k},\omega)=\pi\sum_{\alpha=\pm}
\biggl(1-\frac{\alpha\Delta_{\sigma\xi}}{2|\epsilon_{\sigma\xi}|}\biggr)
\delta(\omega+\alpha|\epsilon_{\sigma\xi}|),
\end{equation}
\begin{equation}\label{eq:A22}
A^{\sigma\xi}_{22}(\bm{k},\omega)=\pi\sum_{\alpha=\pm}
\biggl(1+\frac{\alpha\Delta_{\sigma\xi}}{2|\epsilon_{\sigma\xi}|}\biggr)
\delta(\omega+\alpha|\epsilon_{\sigma\xi}|),
\end{equation}
and $\epsilon_{\sigma\xi}$ is given by Eq.~\eqref{eq:eigen}.  From this we are able to obtain
analytically, the real part 
of the longitudinal
conductivity and its Kramers-Kronig-related imaginary part. These are given as
\begin{align}
\text{Re}\sigma^{\sigma\xi}_{xx}(\Omega)
=&\sigma_0\biggl\{\frac{4\mu^2-\Delta_{\sigma\xi}^2}{4|\mu|}\delta(\Omega)\Theta(2|\mu|-|\Delta_{\sigma\xi}|)\notag\\
&+\frac{1}{4}\biggl[1+\biggl(\frac{\Delta_{\sigma\xi}}{\Omega}\biggr)^2\biggr]
\Theta(\Omega-\Omega_c)\biggr\},\\
\text{Im}\sigma^{\sigma\xi}_{xx}(\Omega)
=&\sigma_0\biggl\{\frac{4\mu^2-\Delta_{\sigma\xi}^2}{4\pi\Omega|\mu|}+\frac{1}{4\pi}\biggl[1+\frac{\Delta_{\sigma\xi}^2}{\Omega^2}\biggr]f(\Omega)\notag\\
&+\frac{\Delta^2_{\sigma\xi}}{2\pi\Omega\Omega_c}\biggr\},
\end{align}
where $\Omega_c={\rm max}(2|\mu|,|\Delta_{\sigma\xi}|)$,
$f(\Omega)=\ln[|\Omega+\Omega_c|/|\Omega-\Omega_c|]$, and
 $\sigma_0=e^2/4\hbar$. 
Likewise, the imaginary part of the zero temperature transverse or dynamical Hall conductivity for a given spin and valley is given by
\begin{align}
\text{Im}&\sigma^{\sigma\xi}_{xy}(\Omega)
=\frac{e^2}{2\Omega}\int_{|\mu|-\Omega}^{|\mu|}\frac{d\omega}{2\pi}\int\frac{d^2k}{(2\pi)^2}v^2
\notag\\
&\times\bigg[
A^{\sigma\xi}_{11}(\bm{k},\omega)A^{\sigma\xi}_{22}(\bm{k},\omega+\Omega)
-A^{\sigma\xi}_{22}(\bm{k},\omega)A^{\sigma\xi}_{11}(\bm{k},\omega+\Omega)
\bigg],
\end{align}
 where $A_{11}(\bm{k},\omega)$ and $A_{22}(\bm{k},\omega)$ are given by Eqs.~\eqref{eq:A11} and \eqref{eq:A22}, respectively.  This can be solved analytically for the imaginary part and the Kramers-Kronig-related real part.  These are
\begin{align}
\text{Im}\sigma^{\sigma\xi}_{xy}(\Omega)
&=-\xi\sigma_0\frac{\Delta_{\sigma\xi}}{2\Omega}\Theta(\Omega-\Omega_c),\\
\text{Re}\sigma^{\sigma\xi}_{xy}(\Omega)
&=\xi\sigma_0\frac{\Delta_{\sigma\xi}}{2\pi\Omega}f(\Omega).
\end{align}
Similar analytic forms have been found in other works for other 
systems.\cite{Gusynin:2006,Tse:2010}
With these formulas in hand, we can now discuss the longitudinal conductivity
and the conductivity for circular polarization associated with the different
valley and spin degrees of freedom. Our aim is to identify signatures of 
of the VSPM state and the TI to BI transition.

\begin{figure}
  \begin{center}
  \includegraphics[width=0.9\linewidth]{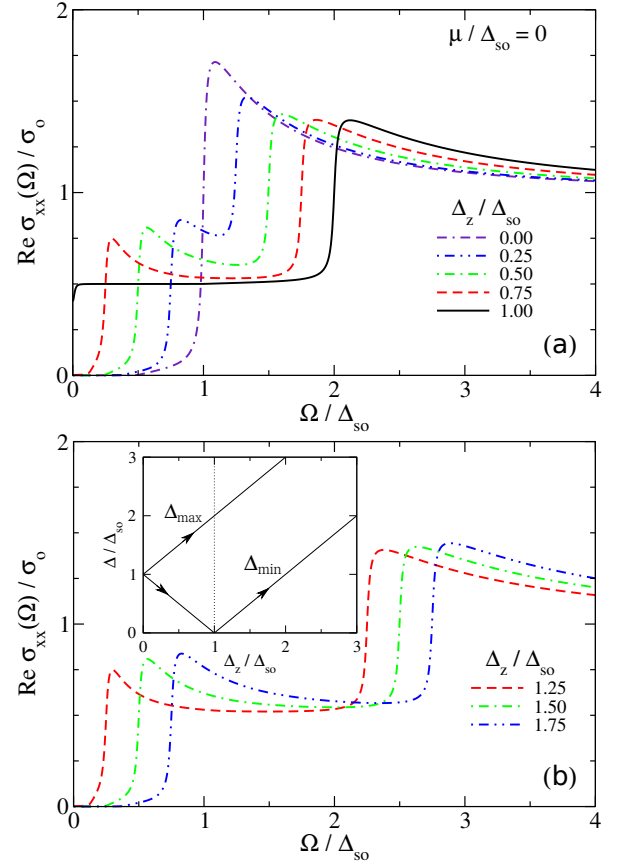}
  \end{center}
  \caption{\label{fig4}(Color online) The real part
of the dynamical conductivity $\text{Re}\sigma_{xx}(\Omega)/\sigma_0$ 
for $\mu=0$ and varying (a) $\Delta_z/\Delta_{\rm SO}<1$ 
and (b) $\Delta_z/\Delta_{\rm SO}>1$. Inset: A plot of the minimum and
maximum band gaps, reproduced from Fig.~\ref{fig3}, showing the
evolution of the absorption peaks with $\Delta_z$. 
}
\end{figure}

\begin{figure}
  \begin{center}
  \includegraphics[width=0.9\linewidth]{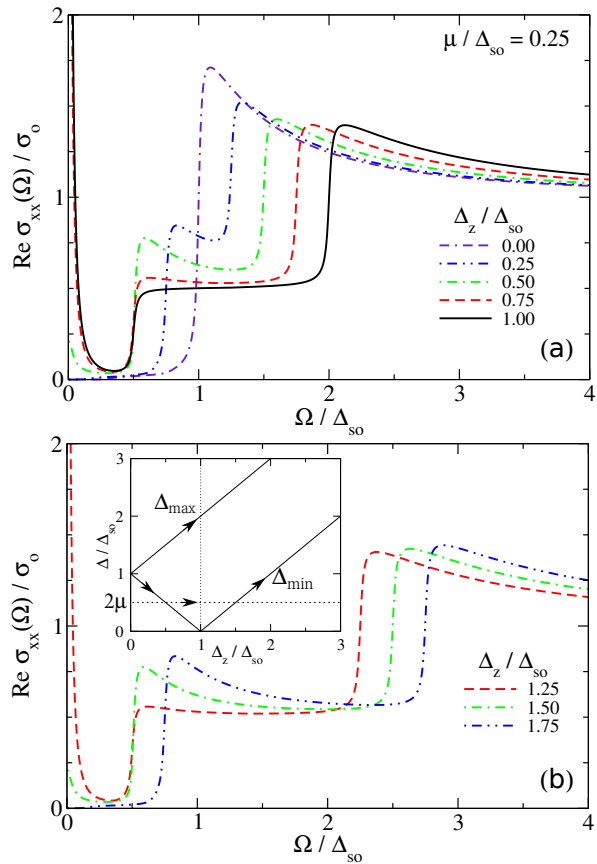}
  \end{center}
  \caption{\label{fig5}(Color online) The real part
of the dynamical conductivity $\text{Re}\sigma_{xx}(\Omega)/\sigma_0$ 
for $\mu/\Delta_{\rm SO}=0.25$ and varying (a) $\Delta_z/\Delta_{\rm SO}<1$ 
and (b) $\Delta_z/\Delta_{\rm SO}>1$.
Inset: A plot of the minimum and
maximum band gaps, reproduced from Fig.~\ref{fig3}, showing the
evolution of the absorption peaks with $\Delta_z$ but with a lower
bound of $2\mu$.  
}
\end{figure}

\section{Longitudinal optical response}

In this section, we discuss the results for the real part of
the frequency-dependent longitudinal conductivity. 
For our calculations shown in the figures, we have
chosen to evaluate Eqs.~\eqref{eq:totcondxx} and \eqref{eq:numcondxx},
where we have replaced the delta functions in the spectral functions
with Lorentzians according to $\delta(x)\to (\eta/\pi)/(x^2+\eta^2)$
with $\eta$, a broadening parameter taken to be $\eta=0.01\Delta_{\rm SO}$.
(Note that the Kramers-Kronig related piece of the delta function
in ${\rm Im}\sigma_{xx}^{\sigma\xi}(\Omega)$
will have to be replaced with that for the Lorentzian form.)\cite{Tabert:2012} 
This mimics
impurity scattering in the system, with the transport 
scattering rate $1/\tau_{imp}=2\eta$.
This is done to provide more
realistic looking curves, however, the
analytical formulas above work very well if the delta function is replaced
with a Lorentzian with halfwidth 
associated with the transport impurity scattering 
rate, instead of $\eta$ which is  the quasiparticle scattering rate in
the spectral function.\cite{Nicol:2008,Tabert:2012}

For the case of
charge neutrality, where $\mu=0$, we have results shown in Fig~\ref{fig4}.
The conductivity is entirely based on interband transitions. For $\Delta_z=0$,
we have only spin orbit coupling, the bands are degenerate in spin with a 
single gap, and there is one absorption edge at $\Delta_{\rm SO}$ which shows
a peak before decaying to the universal background expected for a monolayer
graphene-like system. With finite $\Delta_z<\Delta_{\rm SO}$ (the TI regime),
two edges appear due to the spin-split bands giving rise to two gaps as
discussed in Fig.~\ref{fig3}. (Note that transitions can only
occur between bands with the same spin index.) 
The absorption edges track the 
gaps shown in Fig.~\ref{fig3}
bottom frame (and reproduced here in the inset for positive $\Delta_z$) such
that the edges move oppositely in frequency - a signature feature of the TI 
region. At the VSPM point ($\Delta_z/\Delta_{\rm SO}=1$), the VSPM gives the
famous Dirac massless fermion conductivity, which is flat, but the amplitude is
only one half that of $\sigma_0$ which is different from the case of graphene
where it is $\sigma_0$. At the energy scale of $2\Delta_{\rm SO}$ there is
now an absorption edge leading to a peak structure, as before, with the
onset of  the interband
transitions between the gapped bands. This curve is a signature for 
the VSPM state. For the region for the BI ($\Delta_z>\Delta_{\rm SO}$), two
interband edges occur, the peaks of which both move outward according to the
sum and difference of $\Delta_z$ and $\Delta_{\rm SO}$, but with a constant
separation in energy of $2\Delta_{\rm SO}$. The behavior of the absorption
edge peaks from the TI to BI regimes tracks the arrows of the inset,
i.e., as $E_z$ increases from zero, 
the two peaks move oppositely in the TI region and 
then both move to higher frequency in the BI region.
The plots shown here are for the total longitudinal conductivity. For the
longitudinal conductivity about one $K$ point, the same frequency dependence
remains and one merely has half the magnitude, although the
lower frequency segment would represent the response of charges with one
particular spin orientation and the larger energies are associated with the
superposition of the response of both spin orientations. 
It is clear that from
the frequency dependence, one could identify the spin orbit energy gap value
by tracking the behavior with perpendicular electric field. The extrapolation
of the minimum gap to zero as a function of $E_z$ 
identifies the critical electric field that 
produces the VSPM, and the identification of the two regimes, TI versus BI, is
found by looking for co-moving or counter-moving peaks.

\begin{figure}
  \begin{center}
  \includegraphics[width=0.9\linewidth]{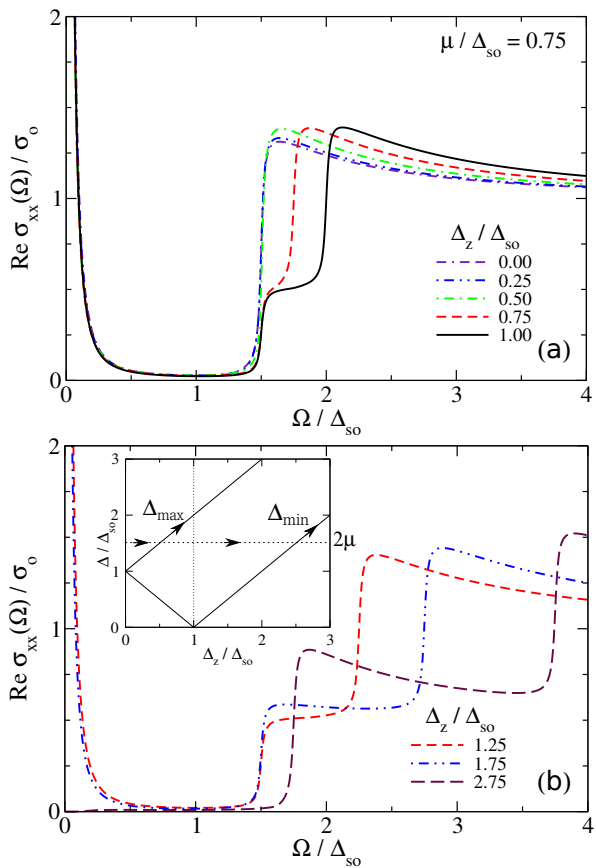}
  \end{center}
  \caption{\label{fig6}(Color online) The real part
of the dynamical conductivity $\text{Re}\sigma_{xx}(\Omega)/\sigma_0$ 
for $\mu/\Delta_{\rm SO}=0.75$ and varying (a) $\Delta_z/\Delta_{\rm SO}<1$ 
and (b) $\Delta_z/\Delta_{\rm SO}>1$. Inset: A plot of the minimum and
maximum band gaps, reproduced from Fig.~\ref{fig3}, showing the
evolution of the absorption peaks with $\Delta_z$. For small $\Delta_z$
there is a lower bound of $2\mu$ before one peak emerges and then eventually
a second peak. 
}
\end{figure}

\begin{figure}
  \begin{center}
  \includegraphics[width=0.9\linewidth]{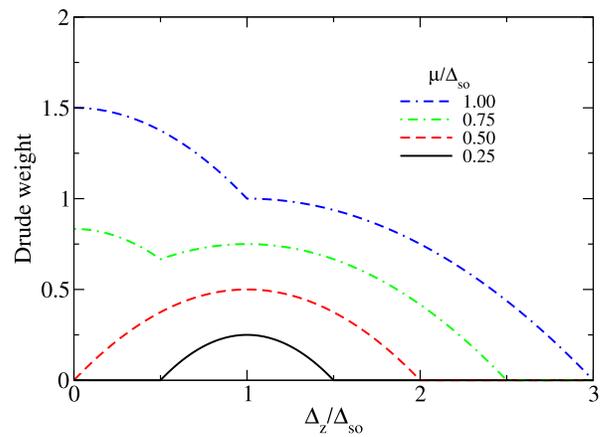}
  \end{center}
  \caption{\label{newfig}(Color online) The Drude weight as a function
of $\Delta_z/\Delta_{\rm SO}$ for varying $\mu/\Delta_{\rm SO}$.
}
\end{figure}

Of course, some things change in the doped case. Primarily, the lowest
energy for absorption at a given electric field strength
is set by $2\mu$ or $\Delta$, whichever is greater. This is most easily
illustrated with Figs.~\ref{fig5} and \ref{fig6} where twice the chemical
potential is taken to be finite but less than $\Delta_{\rm SO}$ in the
former case and greater than $\Delta_{\rm SO}$ in the latter one. The $2\mu$
Pauli-blocking cutoff is well-known in graphene where vertical transitions
from an occupied state to an unoccupied state cannot occur in the band
structure for photon absorption frequency $\Omega<2|\mu|$. In Figs.~\ref{fig5}
and \ref{fig6}, we see that the same pattern from the $\mu=0$ case is occurring
with the exception that there is a lower energy cutoff based on the inset
figures illustrating the progression of the energies  of the absorption edges 
with varying $\Delta_z$ (following the arrows in the inset). 
For $\mu/\Delta_{\rm SO}=0.25$, $2\mu=0.5\Delta_{\rm SO}$ 
and the lower absorption edge decreases until $\Omega=2\mu$ where it remains stationary 
until it starts to increase in frequency at $|\Delta_z-\Delta_{\rm SO}|>2\mu$.
Note that an intraband Drude conductivity also appears during the range of
$\Delta_z$ where the
lowest absorption edge remains at $2\mu$, {\it i.e.} $2\mu>|\Delta_z-\Delta_{\rm SO}|$ or $2\mu>\Delta_{\rm min}$. At this point the chemical potential 
is in the band and consequently intraband absorption processes can occur.
This leads to an interesting observation that varying electric field at finite
doping could provide a mechanism to switch on and off the Drude response 
at low frequency due to the Fermi level falling outside and inside the band gap
as tuned by $\Delta_z$. 
This is illustrated in Fig.~\ref{newfig} where we plot
the Drude weight (the amount of spectral weight under the delta function
of ${\rm Re} \sigma_{xx}(\Omega)/\sigma_0$ 
for $\Omega=0^+$). Examining the black curve with $\mu/\Delta_{\rm SO}=0.25$,
$2\mu$ is less than the minimum gap $\Delta_{\rm min}=|\Delta_{\rm SO}-\Delta_z|$
for $\Delta_z/\Delta_{\rm SO}<0.5$ at which point only interband transitions
can occur. For increased $\Delta_z$, however, the minimum gap decreases and the
Fermi level is in the band and an intraband Drude absorption occurs. The maximum
occurs at the point at which the VSPM is reached and then the Drude weight decreases as the minimum gap once again increases with $\Delta_z$ in the BI regime. When the electric field strength becomes strong enough to increase the minimum gap value further, the Fermi level will once again fall in the gap rather than in
the spin split band and the Drude intraband absorption will disappear (as it does here at $\Delta_z=1.5\Delta_{\rm SO}$).
Thus, for
$2\mu<\Delta_{\rm SO}$, the Drude weight appears
within the region where the minimum gap drops below $2\mu$ 
(as shown in the inset of Fig.~\ref{fig5}). For $2\mu>\Delta_{\rm SO}$, 
the Drude weight is always present until
the minimum gap overcomes $2\mu$ (refer to the inset of Fig.~\ref{fig6}).
The kink in the curve occurs when $2\mu$ is equal to the maximum gap.
Moreover, if the Fermi level is between the spin split bands, the Drude weight will show an increase in the TI region as $\Delta_z$ is increased
and then decrease again as the system is tuned into the BI region. 

\section{Circular polarization}

Circularly polarized light has been suggested as a mechanism for
seeing the valley-spin coupling in silicene\cite{Ezawa:2012opt} and 
in group VI-dichalcogenides\cite{Xiao:2012,Cao:2012}. 
In light of this suggestion, we discuss the full
frequency dependence of the optical response of silicene
to  circularly polarized light. Our discussion is somewhat different from
that of other references\cite{Xiao:2012,Cao:2012,Ezawa:2012opt} where the focus 
has been on the matrix elements evaluated in momentum space for specific frequencies. 
Here we look
at the broadband response. 
For finite frequency optical response  in the circular polarization basis,
the conductivity is given as $\sigma_{xx}(\Omega)\pm i\sigma_{xy}(\Omega)$ for 
righthanded (${+}$) and lefthanded (${-}$) circular polarization.\cite{Pound:2012}
The real part of this function provides  the absorptive part of 
the optical response: 
\begin{equation}
\sigma_\pm(\Omega)=
{\rm Re}\sigma_{xx}(\Omega)\mp{\rm Im}\sigma_{xy}(\Omega),
\end{equation}
which can be obtained from
our previous formulas. 
This form refers to the total absorptive response to circularly polarized light
which is written in terms  of the absorptive parts of the longitudinal
and transverse conductivities which have been
 summed over valley and spin indices.
It will facilitate our discussion to show the components
which are summed over spin but plotted separately for each
valley index, ie., 
\begin{equation}
\sigma^{\xi}_\pm(\Omega)=\sum_{\sigma}[
{\rm Re}\sigma^{\sigma\xi}_{xx}(\Omega)\mp{\rm Im}\sigma^{\sigma\xi}_{xy}(\Omega)],\label{eq:sigpmxi}
\end{equation}
and  we decompose further to show the individual components based on 
valley and spin (the quantity in the square brackets of Eq.~\eqref{eq:sigpmxi}).

\begin{figure}
  \begin{center}
  \includegraphics[width=1.0\linewidth]{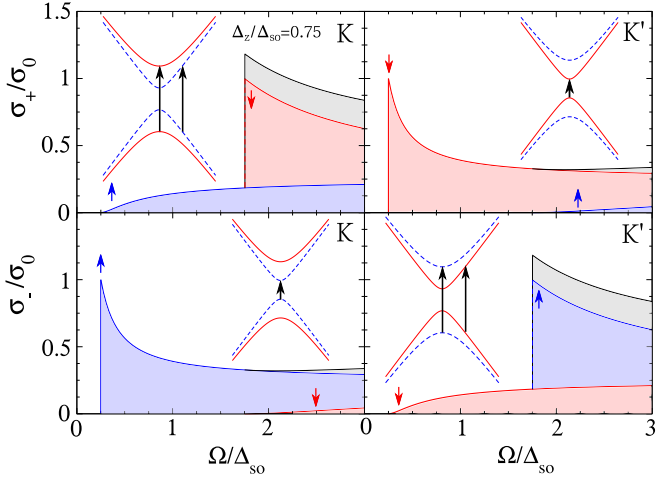}
  \end{center}
  \caption{(Color online) \label{fig7}
The optical response to circularly polarized light for a
TI with spin-split bands. The response is
shown separately for each
$K$ and $K'$ points for righthanded
polarization, $\sigma_{+}(\Omega)/\sigma_0$,
and  lefthanded polarization, $\sigma_{-}(\Omega)/\sigma_0$.
Here, $\Delta_z=0.75\Delta_{\rm SO}$ and $\mu=0$.
The blue shading is for the up-spin band, the red for the down spins,
and the total response in one valley is the envelope of these curves
following the black curve (sum of the red and blue curves)
at higher frequency. The insets show the primary
transitions occurring at the peak absorption frequency.
}
\end{figure}

\begin{figure}
  \begin{center}
  \includegraphics[width=1.0\linewidth]{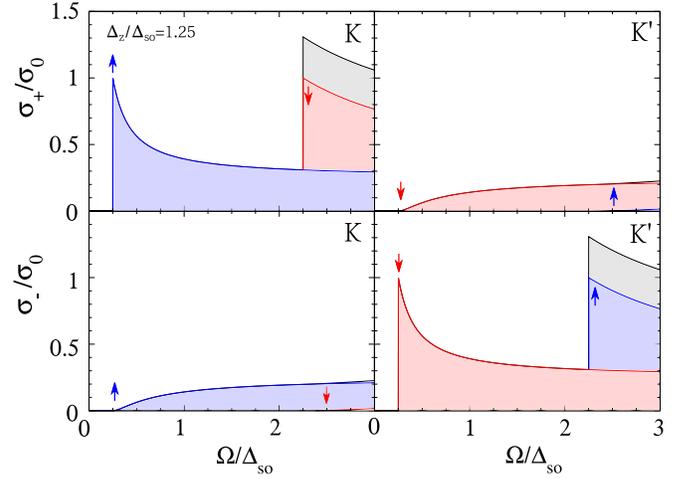}
  \end{center}
  \caption{(Color online) \label{fig8}
The optical response to circularly polarized light 
for a BI with spin-split bands, at the
$K$ and $K'$ points for righthanded
polarization, $\sigma_{+}(\Omega)/\sigma_0$
and  lefthanded polarization, $\sigma_{-}(\Omega)/\sigma_0$.
Here, $\Delta_z=1.25\Delta_{\rm SO}$ and $\mu=0$.
The notation is the same as for Fig.~\ref{fig7}.
}
\end{figure}

To begin with, we show in Figs.~\ref{fig7} and \ref{fig8} 
plots to illustrate the
essence of our results. We have chosen for these plots to use the
analytical formulas in order to simplify and sharpen the structure and to allow
us to make comparison with other figures later on for a discussion about the
effect of broadening which we have not had up till now. 
In Fig.~\ref{fig7}, we show the case
of $\Delta_z/\Delta_{\rm SO}=0.75$ which is the TI regime
and in Fig.~\ref{fig8} we take $\Delta_z/\Delta_{\rm SO}=1.25$
which is in the BI regime. In the four frames,
we plot the response to circularly polarized light for each combination
of $K$ and $K'$ point and $+/-$ polarization. The blue area indicates
the spin up component and the red region, the spin down (as indicated by
the arrows). 

Considering first Fig.~\ref{fig7},
it is quite apparent that with positive polarization and a 
frequency associated with the minimum gap $\Omega_1=|\Delta_{\rm SO}-\Delta_z|=0.25\Delta_{\rm SO}$, the
optical response is dominated by charge carriers with down spins residing
in the $K'$ valley, whereas, for negative polarization, at the
same energy, the response is
due to charge carriers with up spins in the $K$ valley. Hence, circularly
polarized light provides a means to populate valley-specific carriers with
a particular spin orientation. The frequency required is that corresponding
to the minimum gap in the band structure and indicated by an arrow
 on the band structure insets of
the upper right frame and lower left frame of Fig.~\ref{fig7}.
Such an idea has been discussed previously
for silicene by Ezawa\cite{Ezawa:2012opt}
and by others for the dichalcogenides.\cite{Xiao:2012,Cao:2012} 
It was also argued in those works 
that using a photon energy associated with the maximum gap  would connect
the two outer bands, thus allowing for absorption at that frequency which
would be associated with the other spin orientation for each valley. That is,
for $\Omega_2=\Delta_{\rm SO}+\Delta_z=1.75\Delta_{\rm SO}$ shown here,
the dominate response is to see the down spin charge carriers in the $K$
valley and the up spin ones for the $K'$ point. However, as is clear
from these four frames, the circularly polarized response at $\Omega_2$
will not be purely of one spin orientation, nor when the response of the
two valleys are added for a particular polarization, 
will it  be the response of just one valley at that
frequency. As shown in the inset band structure pictures for the
upper left and lower right plots, even though $\Omega_2$ will connect
the two outer bands at $k=0$ to give a peak in the absorption, 
it will also connect
states from the inner two bands at finite $k$, 
and the total absorption is the sum of
the two (the black curve). 
Thus, the lowest frequency $\Omega_1$ is robust
for producing valley-spin polarized charge carriers but the upper frequencies
will be contaminated and not result in a pure spin-valley polarized response. 
So we conclude that
using photons with energy $\Omega_1$ will produce ($K$,$\uparrow$) and 
($K'$,$\downarrow$) 
in this case, given negative and positive circular
polarization, respectively.
The question might then be asked as to whether one could produce the other
combinations in an uncontaminated way. If one returns to the bottom 
frame of Fig.~\ref{fig3}, it can be noted that if $\Delta_z\propto E_z$ 
changes sign, this reverses the spin labels on the bands at the $K$ valley
(indicated by the dashed and solid lines) and likewise at the $K'$ valley
and so, at least from a theoretical point of view, 
($K$,$\downarrow$) and ($K'$,$\uparrow$) 
could also be excited at $\Omega_1$ with oppositely polarized light
if one could reverse the electric field without changing anything else.

In the BI regime, shown in Fig.~\ref{fig8}, the essential difference
is that the lower band spin response is switched between the two 
polarizations for a fixed 
$K$ point and so a double peak structure
appears in the positive polarization response of the $K$ valley
and very little occurs in the negative polarization at the same point
(with the results reversed at the $K'$ valley). Such behavior
was also noted by Ezawa 
and is due to band inversion in the BI relative to the TI.\cite{Ezawa:2012opt}
Thus, one finds that once again, the lowest energy defined as $\Omega_1$
as before, will reflect a single spin state, but now associated with the
opposite polarization from the case of the TI. For the total 
response to polarized light as a function of frequency, in the case of a BI,
 $\sigma_+$ probes mainly the $K$ point and $\sigma_-$,
the $K'$ one, and the two peaks in the frequency dependence
are associated with
different spin orientation. Whereas for the TI, $\sigma_+$ excites primarily down spins and $\sigma_-$,
up spins, and it is two peaks in the frequency dependence 
that resolve the valley-dependence.

\begin{figure}
  \begin{center}
  \includegraphics[width=0.9\linewidth]{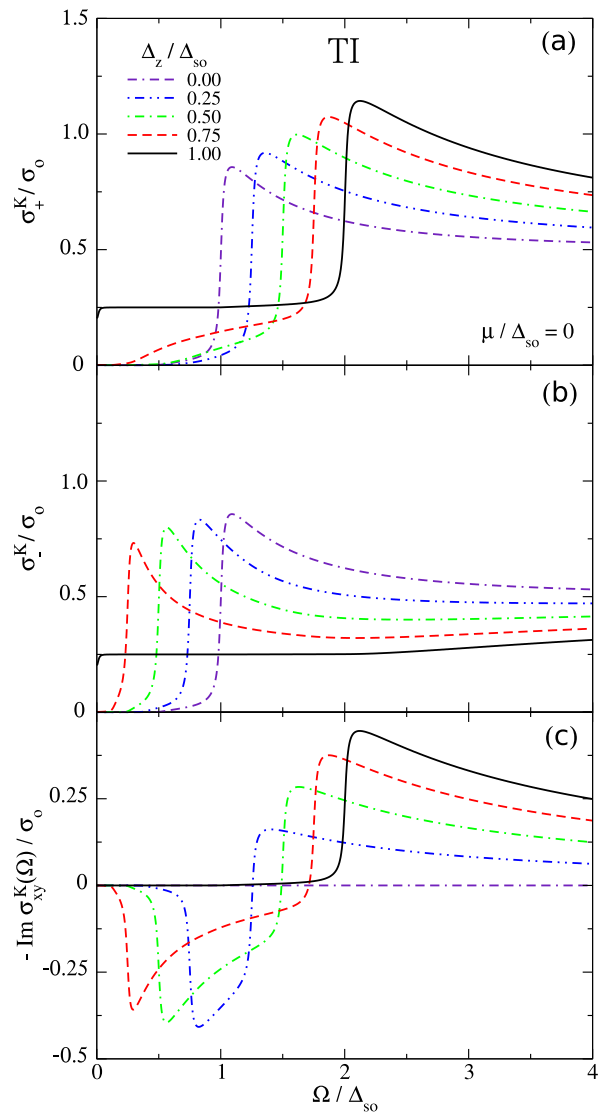}
  \end{center}
  \caption{(Color online) \label{fig9}
The optical response to circularly polarized light at the
$K$ point in the
TI phase with $\mu=0$ (a) Conductivity for righthanded
polarization, $\sigma_{+}(\Omega)/\sigma_0$. (b) Conductivity
for lefthanded polarization, $\sigma_{-}(\Omega)/\sigma_0$.
(c) The transverse Hall conductivity $-{\rm Im}\sigma_{xy}(\Omega)$
which is also $(\sigma_{-}-\sigma_+)/2$. 
The peaks in (b) are associated with spin up charge carriers and those in
(a) are predominantly due to spin down charge carriers.
}
\end{figure}

\begin{figure}
  \begin{center}
  \includegraphics[width=0.9\linewidth]{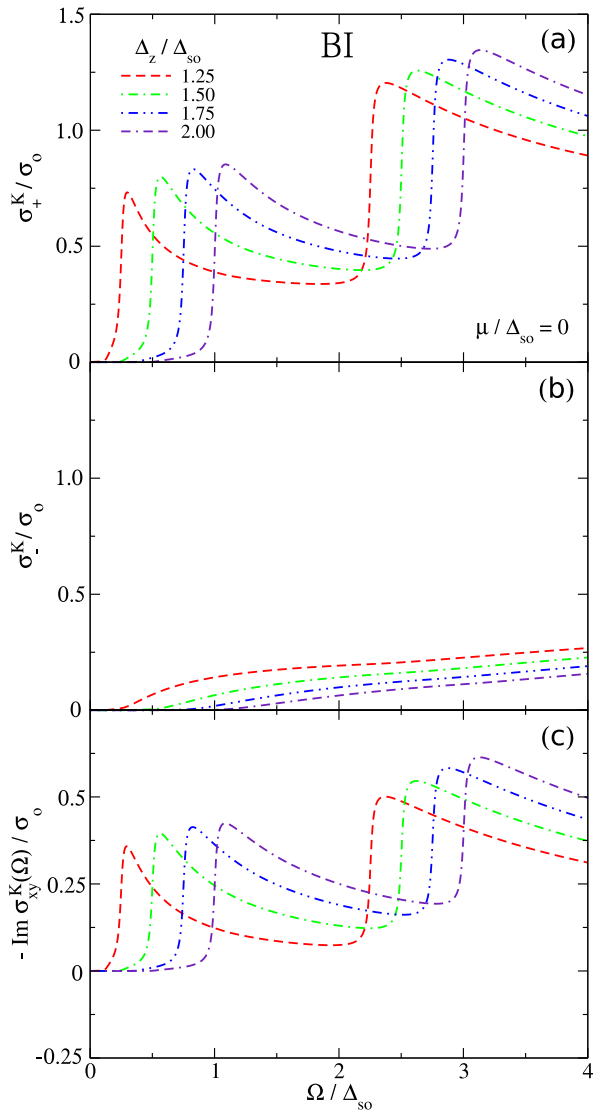}
  \end{center}
  \caption{(Color online) \label{fig10} As for Fig.~\ref{fig10}
but for the BI phase. (a) $\sigma_{+}(\Omega)/\sigma_0$. (b) 
$\sigma_{-}(\Omega)/\sigma_0$. (c) ${-\rm Im}\sigma_{xy}(\Omega)=(\sigma_{-}-\sigma_{+})/2$. The peaks at lower energy are associated with spin up
charge carriers and those at higher with predominantly spin down ones.
}
\end{figure}

Now we comment on the variation of these curves with $\Delta_z$. Given that
the $K'$ valley is always the same but 
reversed with respect to the $K$ valley, as we have discussed, we will
only show the $K$ valley results in the following. 
In Figs.~\ref{fig9} and \ref{fig10}, we show the response for $\mu=0$ for
circularly polarized light $\sigma_+(\Omega)$,
$\sigma_-(\Omega)$, and the imaginary part of the Hall conductivity ${\rm Im}\sigma_{xy}(\Omega)$ as we vary $\Delta_z$. Note that 
$-{\rm Im}\sigma_{xy}(\Omega)=[\sigma_{+}(\Omega)-\sigma_{-}(\Omega)]/2$
and is therefore a measure of circular dichroism.
These figures are only for the one $K$ point. For the
$K'$ point, the curves are the same except reversed 
in polarization label and spin orientation information (see Figs.~\ref{fig7}
and \ref{fig8}) and reversed in sign in the case of $-{\rm Im}\sigma_{xy}(\Omega)$.
The cases for the TI and VSPM are shown in Fig.~\ref{fig9}
and the BI, in Fig.~\ref{fig10}. The first thing to note is that
compared to Figs.~\ref{fig7} and \ref{fig8}, the peaks are suppressed
and broadened due the impurity scattering included in these curves,
Note, for instance, that the lowest peaks in Fig.~\ref{fig9}(b)
are not at a value of 1 as seen in Fig.~\ref{fig7}, but are reduced by 
a significant amount. 
This is due to the broadening and the
reduction is stronger when the energy scale of the peak edge
is smaller and therefore closer to the energy scale of the broadening 
parameter. As a result, these curves present a more realistic sequence
in
behavior with variation with $E_z$. While we do not identify the
separate spin components as in Figs.~\ref{fig7} and \ref{fig8},
one can see that the overall envelope of the curves in Figs.~\ref{fig7}
and \ref{fig8} ({\it i.e.}, the sum of blue and red to give the black curve)
is the same here and so one may use Figs.~\ref{fig7} and \ref{fig8}
as a guide to identify the spin orientations associated with the
peak structures in Figs.~\ref{fig9} and \ref{fig10}. Once
again for the TI case, the pattern remains that the peak seen
in the lefthanded polarization $\sigma_-$ is primarily due to
up spins and that seen in the righthanded polarization $\sigma_+$
is due to down spins superimposed on a background of absorption
due to transitions between  the up-spin bands. In the BI case,
there is only a response in one of the
polarizations (righthanded in this case),
 however the lower peak is spin-selected
for up spins while the upper absorption edge is 
again an admixture of the response of
two spin
species, dominated by down spins. 
The lefthanded polarization shows virtually no response by comparison.
The interesting feature of these curves is that the absorption
edges move with electric field tuning as seen before in the
case of the longitudinal conductivity. As a result, the 
frequency at which the light can excite carriers with a certain
spin orientation in a particular valley
can be tuned ({\it i.e.}, the $\Omega_1$  discussed before depends on
$\Delta_z\propto E_z$).
Following the discussion in our previous section for the longitudinal
conductivity, there will
be  movement of these
peaks in frequency according to the
minimum and maximum gap and they will be suppressed by a $2\mu$ cutoff when
$2\mu$ is greater than the gap energy. Such behavior will allow for
additional flexibility in tuning the frequency at which valley-spin polarized
carriers are excited, and the curves shown here illustrate that the response
may not always be 100\%.
 
If one takes the difference between these
two polarizations, {\it i.e.}, $\sigma^K_+-\sigma^K_-$ to obtain $-2{\rm Im}\sigma^K_{xy}(\Omega)$, a change in
sign is seen for the TI case at the energy of  
 $\Omega_2=\Delta_z+\Delta_{\rm SO}$ but not for the case of the BI,
where it is merely additive.
As we have seen, $\Omega_2$ 
is the point where the dominant response transitions from
 one spin species
to the other. 
Of course, due to the sign change from
one valley to another in ${\rm Im}\sigma^\xi_{xy}$, the total response for
${\rm Im \sigma_{xy}(\Omega)}$ is identically zero. 
We note at this point that the total $\sigma_\pm(\Omega)$, summed over valleys
and spins, will be the same function
as ${\rm Re} \sigma_{xx}(\Omega)$, but
that the character of the absorptive response for circular polarization 
of light will be
associated uniquely with a particular spin and valley index, particularly
at the lowest absorption edge. Consequently, keeping this in mind, 
we do not show results for finite doping but use the longitudinal
case for reference. The primary effect of 
 finite doping where $2\mu>\Omega_1$ is to
 eliminate absorption at lower frequencies and induce
a Drude response. However, the Drude response will not show 
spin-valley polarization under circularly polarized light.

Finally we discuss the VSPM. This is shown as the solid black curve in Fig.~\ref{fig9} for $\Delta_z/\Delta_{\rm SO}=1$. The low frequency response
 of the Dirac-type dispersion is the flat background seen in (a) and (b) 
 associated with up spins  of the $K$ valley. The
magnitude is $\sigma_0/4$ in both cases. 
The result at $K'$ is reversed
such the flat background is due to down spins. As the two $K$ points must
be added for the total $\sigma_\pm$, there is an equal admixture of
spin up and spin down, regardless of the sign of circular polarization and so
one would need a mechanism\cite{Xiao:2007,Yao:2008} 
that would allow for sampling one valley only in
order to see a single spin state associated with the linear bands.
The only variation in response 
occurs at $\Omega>2\Delta_{\rm SO}$ when the upper band due to the opposite
spin species enters and
then a peak appears in $\sigma^K_+$ for down spins, superimposed on the
flat background due to up spins.
In ${-\rm Im\sigma^K_{xy}(\Omega)}$  at the one valley, 
there is only one peak  as opposed to two
for the BI or two with opposite sign for the TI with finite $E_z$.
As a final comment, we wish to note that for $E_z=0$, the bands
are not spin-split and as a result, there is no circular
dichroism or valley-spin polarization.

It is important to reiterate
 that while we show valley-separated curves here, the
total response of a particular polarization is the sum over the two
valleys and hence, the result will simply look like the curves
for ${\rm Re}\sigma_{xx}(\Omega)$ seen in previous figures. The point here is that
the lowest peak in those curves is actually associated with a particular
spin and valley when looked at with circular polarization. 
With other mechanisms\cite{Xiao:2007,Yao:2008} for separating out 
the valleys, the extra information shown here may be accessible.

\section{Summary}

We have calculated the optical response of silicene and related
materials, such as germanene,
 which are predicted to exhibit a tunable band gap due to
an applied perpendicular electric field. Due to the interplay of
spin orbit coupling and the electric field strength, the bands display
spin-valley coupling. Tuning of $E_z$, allows for rich behavior varying
from a topological insulator to a band insulator with a valley 
spin-polarized metal at a critical value in between. We find that in the
longitudinal dynamic conductivity, two peaks marking interband absorption edges
will shift oppositely in frequency with increasing $E_z$ in the TI phase
and in the BI phase, will move in tandem to higher frequency, separated by $2\Delta_{\rm SO}$.
The VSPM exhibits a flat graphene-like  conductivity
at lower frequencies changing to a single peak at $2\Delta_{\rm SO}$ due to
a second spin-split band with a band gap of $2\Delta_{\rm SO}$. 
We suggest that seeing
some of these features or tracking the variation of the peaks with $E_z$
should allow for the identification of these phases and the determination of 
parameters such as $\Delta_{\rm SO}$ and 
the critical electric field that produces the VSPM. 
We have also noted the tunable band gap of this system could
allow for a switching on and off of the Drude or DC response
which may be of interest for technological applications.
With $\Delta_{\rm SO}$
predicted to be about 1.5meV in silicene and 25meV in germanene, this 
spectroscopy should fall in the realm of the THz to far-infrared.
These spectroscopic ranges have successfully probed the dynamic conductivity
of a variety of materials, for example, the first observation of the
far-infrared conductivity
of graphene,\cite{Li:2008} and the 1.4 meV energy gap in superconducting Pb
measured most recently in the THz range.\cite{Mori:2008}  Consequently,
this should be quite feasible once appropriate samples are developed.

Another interesting aspect of this study is the consequences 
for circularly polarized light. We have shown that using circularly
polarized light will resolve the spin and valley degrees of freedom.
For instance in the TI with a perpendicular
electric field yielding spin-split bands, 
the spin up charge carriers at the $K$ valley
can be resolved by lefthanded circularly polarized light
and the spin down ones at $K'$ with righthanded polarization. Moreover,
righthanded/lefthanded polarization probes primarily spin up/down carriers
as a function of frequency.
Tuning
the electric field into the BI phase causes a band inversion such that
the $K$-valley spin-up carriers are now resolved with righthanded
polarization and spin-down ones at $K'$ with the lefthanded case,
and circularly polarized light resolves primarily one valley
as a function of frequency. A reversal of
the perpendicular field direction could also reverse the spin-valley
polarization for these two cases. The range of frequency for
obtaining a single spin orientation in the response is largely restricted
to the region about lowest frequency peak in the interband absorption.
Doping above the minimum gap edge
will simply shift the spectral weight 
to an intraband or Drude component which will not be
spin-valley resolved. At present, no experiments of
this type  exist on
silicene, but a demonstration of tuning the band gap from TI to BI
in such systems would be of great interest. Other materials, namely
the group VI dichalcogenides, are in the BI limit and predictions
for the dynamic conductivity
are being examined for MoS$_2$ by others.\cite{Carbotte:private}
Given the potential for further developments with 2D crystals, 
we anticipate that
the optical properties and band gap tuning will play an important role in the
investigation of new materials and the development of new technologies.
  

\begin{acknowledgments} 
We thank Jules Carbotte for discussions.
This research was supported by the Natural Sciences and
Engineering Research Council of Canada.
\end{acknowledgments}


\bibliographystyle{apsrev4-1}
\bibliography{bib}

\end{document}